\documentclass[12pt]{article}
\usepackage{graphicx}
\usepackage{url}
\usepackage{cite}
\usepackage[margin=1.25 in]{geometry}
\usepackage[colorlinks = true, linkcolor = blue, urlcolor = blue,
      citecolor = blue, anchorcolor = blue]{hyperref}

\newcommand\pubnumber{SLAC-PUB-251124}
\newcommand\pubdate{January 2026}

\def\SLAC{SLAC National Accelerator Laboratory,
    Stanford University, Menlo Park, California 94025 USA}
\def\doeack{\footnote{Work supported by the US Department of Energy,
                     contract DE--AC02--76SF00515.}}

\def\Title#1{\begin{center} {\Large #1 } \end{center}}
\def\Author#1{\begin{center}{ \sc #1} \end{center}}

\def\submit#1{\begin{center}Submitted to {\sl #1} \end{center}}
\newcommand\pubblock{\rightline{\begin{tabular}{l}
        \pubnumber \\ \pubdate \end{tabular}}}
\newenvironment{Abstract}{\begin{quotation} \begin{center}
                       ABSTRACT
     \end{center}\bigskip  }{\end{quotation}}

\def\submit#1{\begin{center} #1\end{center}}
\def\Acknowledgements{\bigskip  \bigskip \begin{center} \begin{large}
             \bf ACKNOWLEDGEMENTS \end{large}\end{center}}

\def\beq{\begin{equation}}
\def\eeq#1{\label{#1}\end{equation}}
\def\eeqn{\end{equation}}

\newenvironment{Eqnarray}%
   {\arraycolsep 0.14em\begin{eqnarray}}{\end{eqnarray}}
\def\beqa{\begin{Eqnarray}}
\def\eeqa#1{\label{#1}\end{Eqnarray}}
\def\eeqan{\end{Eqnarray}}

\def\leqn#1{(\ref{#1})}

\let\bar=\overbar

\def\lsim{\mathrel{\raise.3ex\hbox{$<$\kern-.75em\lower1ex\hbox{$\sim$}}}}
\def\gsim{\mathrel{\raise.3ex\hbox{$>$\kern-.75em\lower1ex\hbox{$\sim$}}}}

\def\L{{\cal L}}

\def\L{{\cal L}}

\def\del{\partial}
\def\Dslash{\not{\hbox{\kern-4pt $D$}}}
\def\dslash{\not{\hbox{\kern-2pt $\del$}}}

\def\Dlr{\mathrel{\raise1.5ex\hbox{$\leftrightarrow$\kern-1em\lower1.5ex\hbox{$D$}}}}

\def\ee{e^+e^-}

\def\msb{{\bar{\scriptsize M \kern -1pt S}}}

\def\drb{{\bar{\scriptsize D \kern -1pt R}}}

\makeatletter
\def\section{\@startsection{section}{0}{\z@}{5.5ex plus .5ex minus
 1.5ex}{2.3ex plus .2ex}{\large\bf}}
\def\subsection{\@startsection{subsection}{1}{\z@}{3.5ex plus .5ex minus
 1.5ex}{1.3ex plus .2ex}{\normalsize\bf}}
\def\subsubsection{\@startsection{subsubsection}{2}{\z@}{-3.5ex plus
-1ex minus  -.2ex}{2.3ex plus .2ex}{\normalsize\sl}}

\renewcommand{\@makecaption}[2]{%
   \vskip 10pt
   \setbox\@tempboxa\hbox{\small #1: #2}
   \ifdim \wd\@tempboxa >\hsize     
       \small #1: #2\par          
     \else                        
       \hbox to\hsize{\hfil\box\@tempboxa\hfil}
   \fi}

\makeatother

\begin{document}
\begin{titlepage}
\pubblock

\vfill
\begin{center} 
\Title{The Future of Higgs Boson Physics}
\vfill
\Author{
  Michael E. Peskin\doeack}
 \medskip
\SLAC 
\end{center}
\vfill
\begin{Abstract}
 In this lecture, I discuss measurements of the properties of the
 Higgs boson and related observables in the era of Higgs factories.
This  highly motivated experimental program is the challenge for the
next generation of particle physicists.
\end{Abstract} 
\vfill
\submit{Invited plenary lecture presented at the \\
  32nd International Symposium on Lepton Photon Interactions at High
  Energies, \\
  Madison, Wisconsin, USA, August 25-29, 2025}

\vfill
\end{titlepage}

\hbox to \hsize{\null}


\tableofcontents

\def\thefootnote{\fnsymbol{footnote}}
\newpage
\setcounter{page}{1}

\setcounter{footnote}{0}

\section{Introduction}

The Higgs boson has always been at the center of the Standard Model of
Particle Physics.   Without it, the physics of the Standard Model
would be impoverished, with no spectroscopy or flavor physics,  the
universe a dilute soup of massless mesons and leptons.   The discovery
of the Higgs boson at the Large Hadron Collider in
2012\cite{ATLASHiggs,CMSHiggs}
gave a crucial verification of this important idea.  Since this
discovery, the LHC experiments ATLAS and CMS have demonstrated, for the
$W$ and $Z$ bosons and for an increasing number of the quarks and leptons,
that the Higgs field  provides at least the major source of their
masses~\cite{ATLAScouplings,CMScouplings}.

These experimental advances ought to change our perspective on the
study of particle physics.   We now see clearly that the major
outstanding questions about particle physics are questions about the
interactions of the 
Higgs boson.  We need to learn more about this particle, and we have
the resources to gain that knowledge.   We will learn much from the
High Luminosity phase of the LHC, but, beyond this, we will be able to
measure the properties of the Higgs boson with high precision from a
future ``Higgs Factory'', an $\ee$ collider operating at energies from
the $Z$ boson resonance to the top quark threshold and above.
This has been recognized in
the latest strategy statements from all of the major world
regions in particle physics.  From the 2020 Update of the European
Strategy for Particle Physics: ``An electron-positron Higgs factor is
the highest priority next collider.''~\cite{ESPP2020}.   From the 2022
report of
the Japan Association of High Energy
Physicists (JAHEP): ``An  $\ee$
Higgs Factory is the most important next large-scale particle physics
facility''\cite{JAHEP}.     From the 2023
US High Energy Physics Advisory Panel 
P5 report: ``The next step is to use electron and positron beams to
construct a Higgs factory, which would allow precision measurements of
the Higgs boson properties and searches for exotic decays, possibly
into dark matter''~\cite{P5}.
In China, the Circular Electron-Positron Collider (CEPC) was 
named the highest priority next project in particle and nuclear
physics  in the
 first round of CAS deliberations~\cite{LouCEPC}, though eventually
 the CAS chose to endorse a less ambitious collider.   Very recently,
 the European Strategy Group has recommended the FCC-ee, a circular
 Higgs factory, as the next flagship CERN project~\cite{CERNESG}. 

In this lecture, I review the 
importance of the measurements expected from Higgs factories  and the
levels of precision that they might achieve.
Section~2 reviews the central role of the Higgs boson that
motivates these measurements.  In Sections~3--5, I discuss Higgs
measurements in three different programs available at Higgs factories.
Section~3 discusses the program at the peak of the
$\ee\to HZ$ cross section at 240-250~GeV in the $\ee$ center of
mass. Section~4 discusses the program of precision measurements at
the $Z$ resonance.  Section~5 discusses measurements above the top
quark threshold.  All three sections touch on the question
of the comparison of linear and circular $\ee$ colliders.    The
European Strategy Group has now recommended the FCC-ee as the next
CERN project, but still it is interesting to summarize what we have
learned about this comparison in the past year of physics discussions.
Section~6 discusses the fascinating technical
opportunities that the Higgs factory physics program will provide.  Section~7 will
give some conclusions.

The discussion in Sections~3--5 relies heavily on the excellent
work reported in the Physics Briefing Book for the 2026 update of the
European Strategy for Particle Physics~\cite{deBlas:2025gyz} and, in
particular, on the new global fit defining the capabilities of 
future colliders.  This is a major update of the global analysis
prepared  for the Snowmass 2022 study~\cite{deBlas:2022ofj}.

The discussions for the update of the European Strategy for Particle
Physics have considered both the goal of  improving our knowledge of
the Higgs boson and
the goal of providing a collider that can reach 10~TeV in the parton
center of mass.  In this paper, I will concentrate on the first of
these aims.  This is the program that needs to be  defined
now --- the physics of precision
Higgs boson measurements available with current accelerator
technologies and the strategies for achieving the goals of this
program.  As I will emphasize in this paper, this is the most
important goal concerning future colliders for the progress of our
field and our commitment to the  current generation of young
physicists.
In parallel, we must study the choices for accelerators reaching 
10~TeV parton collisions.  But today there is no affordable
technology that meets this goal, and it will take decades of
accelerator R\&D to develop and evaluate possible paths.

\section{Importance of studying the Higgs boson with high precision}

Yes, what I said above is correct.  The major outstanding questions
about particle physics are in fact questions about the properties of
the Higgs boson.

The Standard Model (henceforth, SM) has two
distinct pieces.  The first is the gauge Lagrangian
\beq
         \L_{gauge} = \sum_{a = 1,2,3} -{1\over 4}(F^a_{\mu\nu})^2 +
         \sum_f \bar f \ i\Dslash\  f \ ,
         \eeqn
    which describes the motion of gauge bosons, quarks, and leptons
    and their couplings to one another through a beautiful geometrical
    structure.
    This Lagrangian has only three
    parameters, the gauge couplings of $U(1)$, $SU(2)$, and $SU(3)$,
   whose values tempt us to believe that they derive from a single
   grand-unified coupling.   The second is the Higgs Lagrangian,
\beq
         \L_{Higgs} = |D_\mu \Phi|^2 - V(\Phi)  - \sum_{f_L,f_R} 
      Y_{f} \  \bar f_L\cdot \Phi f_R + (h.c.)  ,
         \eeqn
where $V(\Phi)$ is the Higgs field potential and the third term is a
shorthand for the quark and lepton Yukawa couplings.

In a
renormalizable theory, $V(\Phi)$ has 2 parameters, the Higgs mass and
self-coupling; if the SM is an effective theory, it can have many
more.   The Yukawa couplings have a total of 54 parameters, of which
only  16 are observable if the SM is exact.  Neutrino masses are
easily accommodated by adding right-handed neutrinos $\nu_{R}$
with their own
mass matrix and Yukawa interactions. The SM gives us no physical explanation
for any of these couplings or any way of computing them from deeper
principles.
Yet, they contain a huge amount of
physics -- the spectrum of quark and lepton masses, the quark flavor
mixings and CP violation , and, in the model with $\nu_R$, the origin
of neutrino masses.  The SM does not explain the preponderance of
matter over antimatter in the universe, but any explanation would
require CP violation due to
scalar bosons, either the Higgs bosons or their possible heavy
partners.  The SM does not contain cosmic dark matter, and there are many
ad hoc models of dark matter than do not invoke the Higgs sector.  But
all dark matter models with other physical motivations, in particular,
explanations involving  supersymmetry or axions, invoke either the
Higgs boson or its partners.

Above all of these questions, there is the question of why the Higgs
field behaves in its special way:  Why does the Higgs field acquire a
vacuum expectation value and break the symmetry of the SM?  There are
many examples of spontaneous symmetry breaking in physics, from
superconductivity and magnetism in condensed matter systems to chiral
symmetry breaking in QCD.  In each of these cases, there is a
compelling physical explanation for the fact that the system forms an
asymmetric ground state.   The Higgs symmetry breaking is the one
exception.
In the SM, the instability of the symmetrical vacuum state is put in
by hand. Theorists have proposed many models of Higgs symmetry
breaking, including models with supersymmetry, models in which
the Higgs
field is
a composite state, and models in which the Higgs field arises from the
physics of
extra space dimensions.  But at this moment there is no experimental evidence for any
of them.  Importantly,   each of these approaches
 gives a different set of  explanations of all of the phenomena listed in the
previous paragraph.  Without the answer to this question, we cannot
make progress against those major problems of particle physics.

To summarize this situation: ``Higgs isn't everything.  It's the only
thing''\footnote{Apologies to Red Sanders~\cite{RedSanders}.}.

The mystery of the origin of electroweak symmetry breaking is tightly
connected to another often-discussed issue in particle physics, the
``hierarchy problem'' or the ``naturalness'' problem.
These days, with the failure, so
far, to discover new particles at the LHC, it has become fashionable
not to believe in naturalness.  I am very much opposed to this
attitude.  For an extended discussion, please see my recent paper on
this question~\cite{MyHierarchy}.   As a point of emphasis, I 
remind you that the naturalness of the electroweak scale
is the only reason that we expect to see
the breakdown of the SM at the 10~TeV scale, as opposed to a much
higher and probably inaccessible energy scale. Would you  build a 10~TeV
collider at great expense  if there is no expectation of a discovery
at that machine?

\begin{figure}[t]
     \begin{center}
       \includegraphics[width= 0.8\linewidth]{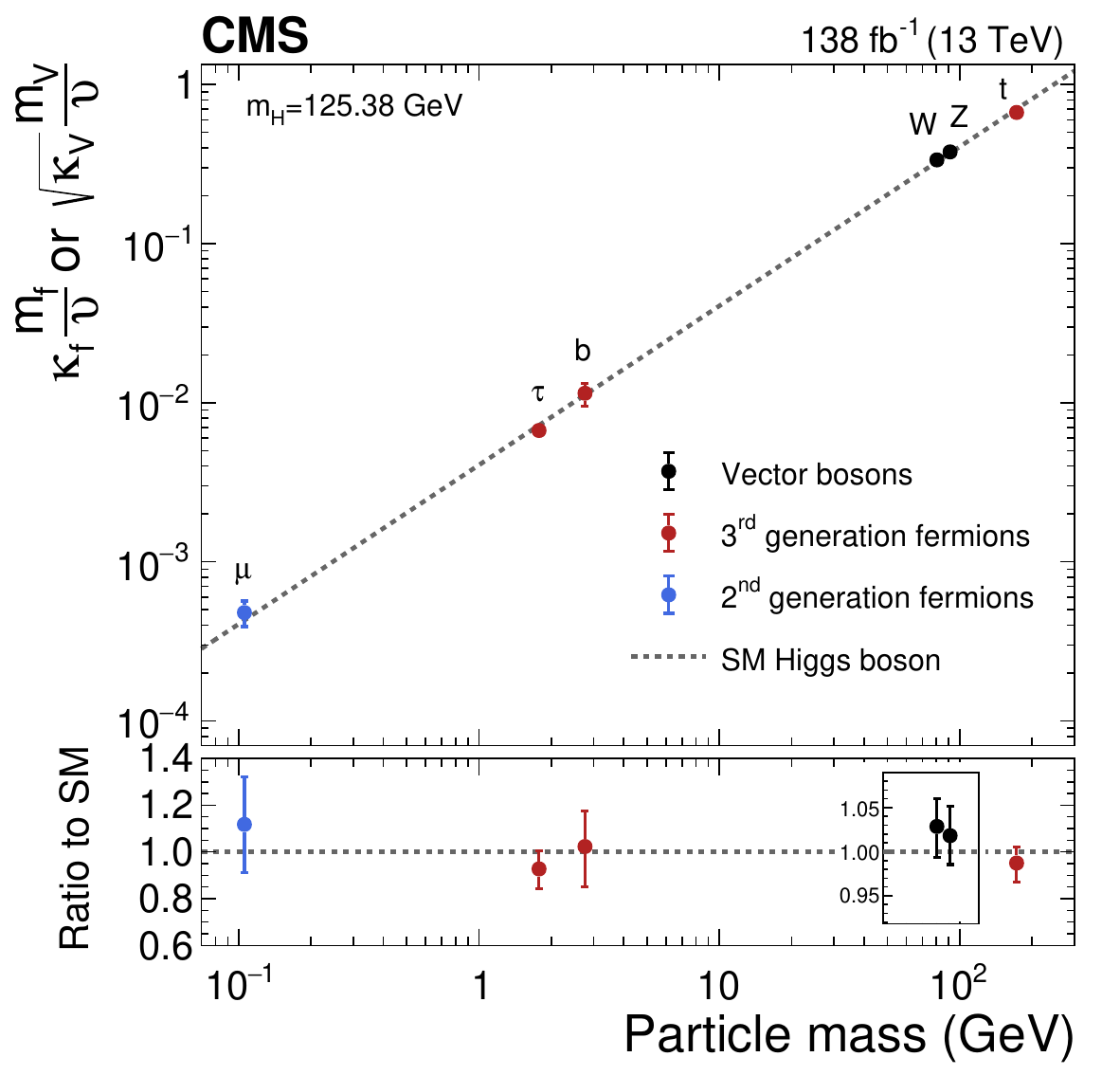}
          \end{center}
\caption{The correspondence between Higgs boson couplings and particle
  mass, according to the Higgs coupling measurements achieved by the
  CMS experiment, from~\cite{CMScouplings}. The lower plot shows the
  residual uncertainties on a linear scale.}
\label{fig:CMSNature}
\end{figure}

There is one more issue concerning the Higgs boson that needs a
comment.   This is the truly impressive job that the ATLAS and CMS
experiments have done in measuring the couplings of the Higgs boson,
using observed Higgs boson decays and also the production of $t\bar t
H$ final states.   The results as of the end of run 2 of the LHC are
collected in Fig.~\ref{fig:CMSNature}, from CMS, and a similar figure
from ATLAS, both published in
Nature~\cite{ATLAScouplings,CMScouplings}.  These  data demonstrate 
that the Higgs boson vacuum expectation value is the
main source of mass for the particles shown.  Many people interpret
these beautiful results as proof that the observed Higgs boson precisely follows
the Standard Model.    To a great extent, this is true, but it is
important to think more carefully. 

The SM relation that, for each particle, the Yukawa coupling is
proportional to the particle mass, can be corrected by the effects of
new particles from a beyond-SM (BSM) theory.   But we should expect that
these corrections are small.  The Yukawa couplings are determined, at
$q^2=0$, from the measured values of the masses.  These couplings, at
$q^2 = m_H^2$, 
can be measured in 
Higgs decays  New
particles of heavy masses $M$ can generate a difference between the SM
prediction and the measured value of the couplings, but this will be
proportional to 
\beq
m_H^2/M^2 \ .
\eeqn
The fact that the LHC has not already discovered these new particles
makes it very likely that their masses are above 1~TeV.   Then the
effects of BSM particles on Higgs couplings are expected to be at the
level of a few percent.   The same conclusion follows from more formal
arguments based on Effective Field Theory or quantum field theory
``decoupling''.   This means that, however impressive the LHC results
are, we are {\it not yet in the game} of observing BSM effects in the Higgs
couplings.

The expected results from the High-Luminosity LHC will
bring us to uncertainties of 2-4\% -- enough to suggest an effect, but
likely not
enough to prove it.

There is a point here that is often ignored in discussions of future
precision measurements:   {\bf The purpose of precision is discovery,
  not just improving the error bars.}   It is possible to make
discoveries with precision, but we must realize that the burden of
proof is much higher than for the discovery of resonances.  The community
will be skeptical about a precision  measurement that deviates from
the SM, and with good reason, considering the recent history of SM
anomalies.
It will be necessary to bring arguments that can overcome
this skepticism. This requires multiple cross-checks on the
measurement and the associated interpretation.
Systematic errors, especially those obtained by human estimation, should be
subdominant.  It is especially important that  the same
deviation should be seen in different settings with different sources
of uncertainty.   I will return to these criteria as I discuss the
discovery opportunities from various stages of the Higgs precision program,

\section{Higgs measurements at threshold}

At the moment, there are a number of Higgs factory proposals under
consideration around the world.
These include circular $\ee$ colliders, the Future Circular
Collider (FCC-ee), 
at CERN~\cite{FCCee,FCCee2,FCCee3,FCCee4}, and the Circular Electron-Positron Collider (CEPC)
in China~\cite{CEPCSnowmass,CEPCTDR}. They also include linear $\ee$ colliders,
such as the ILC in Japan~\cite{ILC}, the Compact Linear Collider
(CLIC) at CERN~\cite{CLIC}, and the relatively newly proposed Linear
  Collider Facility (LCF) at CERN~\cite{LCF}.  All of these, except
  for CEPC, have been studied as candidate future machines for
  the European Strategy for Particle Physics.   For the details of the
  accelerator designs and estimated costs and schedules,
  please follow the references; there is no room for a full discussion
  here.

  \begin{figure}[t]
    \begin{center}
      \includegraphics[width=0.8\linewidth]{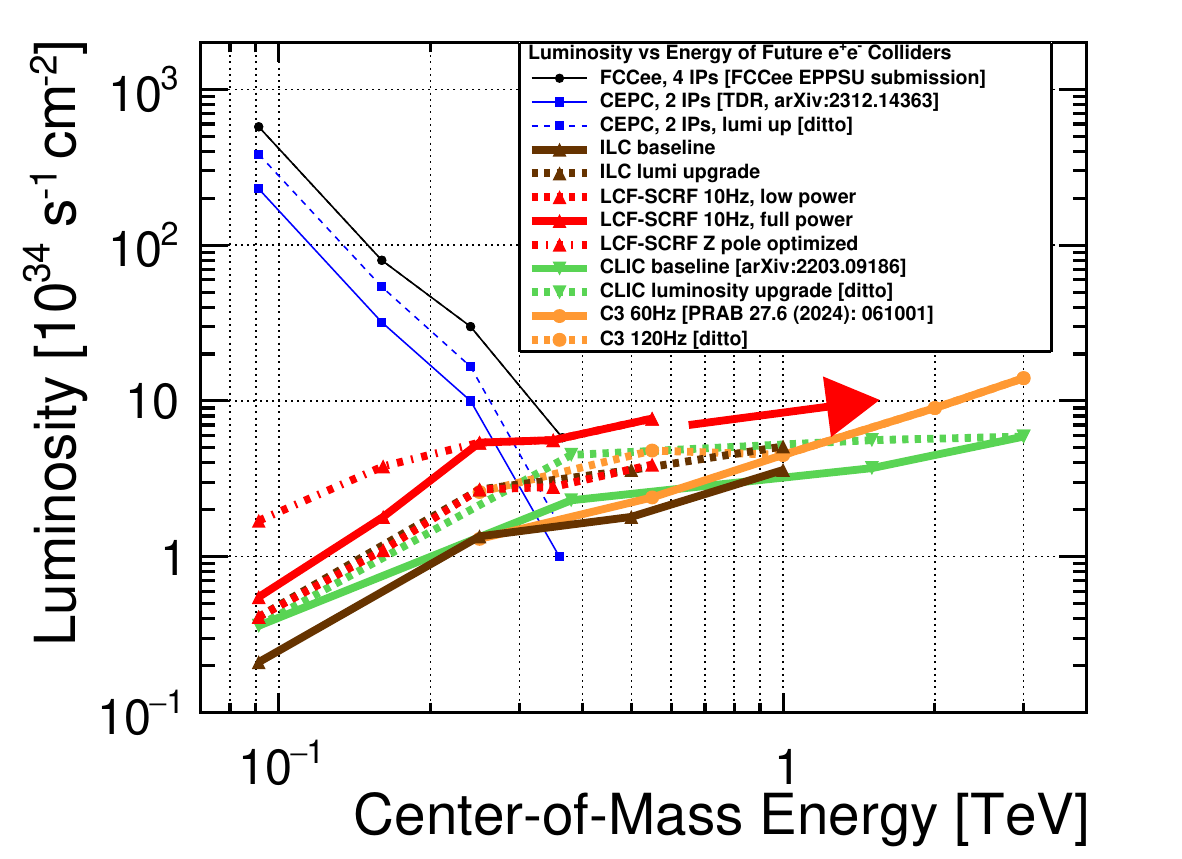}
      \end{center}
\caption{Expected instantaneous luminosities as a function of CM
  energy
  for proposed circiular and linear Higgs factories, from ~\cite{LCVision}.}
\label{fig:Lumis}
\end{figure}

Some crucial difference between the linear and circular designs need
  to be emphasized:   These designs have luminosities that depend in
  very different ways on the CM energy.   For the circular designs,
  the luminosity is limited by synchrotron radiation, leading to a very
  high luminosity at the lowest energies which then falls as
  $E_{CM}^{-4}$ as the CM energy is increased.   Linear colliders
    have luminosities limited by beam-beam  effects, leading to a luminosity
  increasing as $E_{CM}^1$, as is typical for lower-energy
    synchrotrons.
    The expected luminosities for many of the projects cited above are
    shown in Fig.~\ref{fig:Lumis}~\cite{LCVision}.    The crossover is at about
    350~GeV, close to the $t\bar t$ threshold.  The run plans for the
    various projects reflect this dependence.  In particular, the
    luminosity for circular machines is about 3 orders of magnitudes
    higher than for linear machines at the $Z$ pole, but falls too low
    for an interesting physics program at CM energies of 400~GeV and
    above.    Linear colliders can support longitudinal beam
    polarization both for electrons and positrons, an advantage in
    studying the chiral couplings of the  SM.
    We will see how these differences play
    out as we consider the various energy stages of $\ee$ Higgs
    factory operation.

 \begin{figure}
    \begin{center}
      \includegraphics[width= 0.7\linewidth]{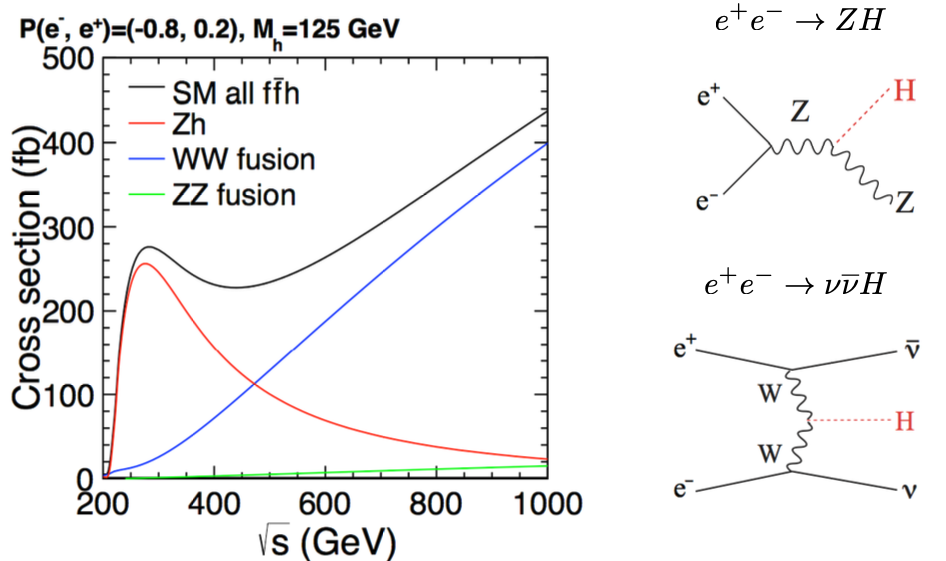}
      \end{center}
\caption{Cross sections for Higgs boson production in $\ee$
  annihilation reactions as a function of CM energy, showing also the
  dominant Feynman diagrams for the reactions $\ee\to ZH$ and $\ee\to
  \nu\bar\nu H$. }
\label{fig:Higgscx}
\end{figure}

  \begin{figure}
    \begin{center}
      \includegraphics[width=0.6\linewidth]{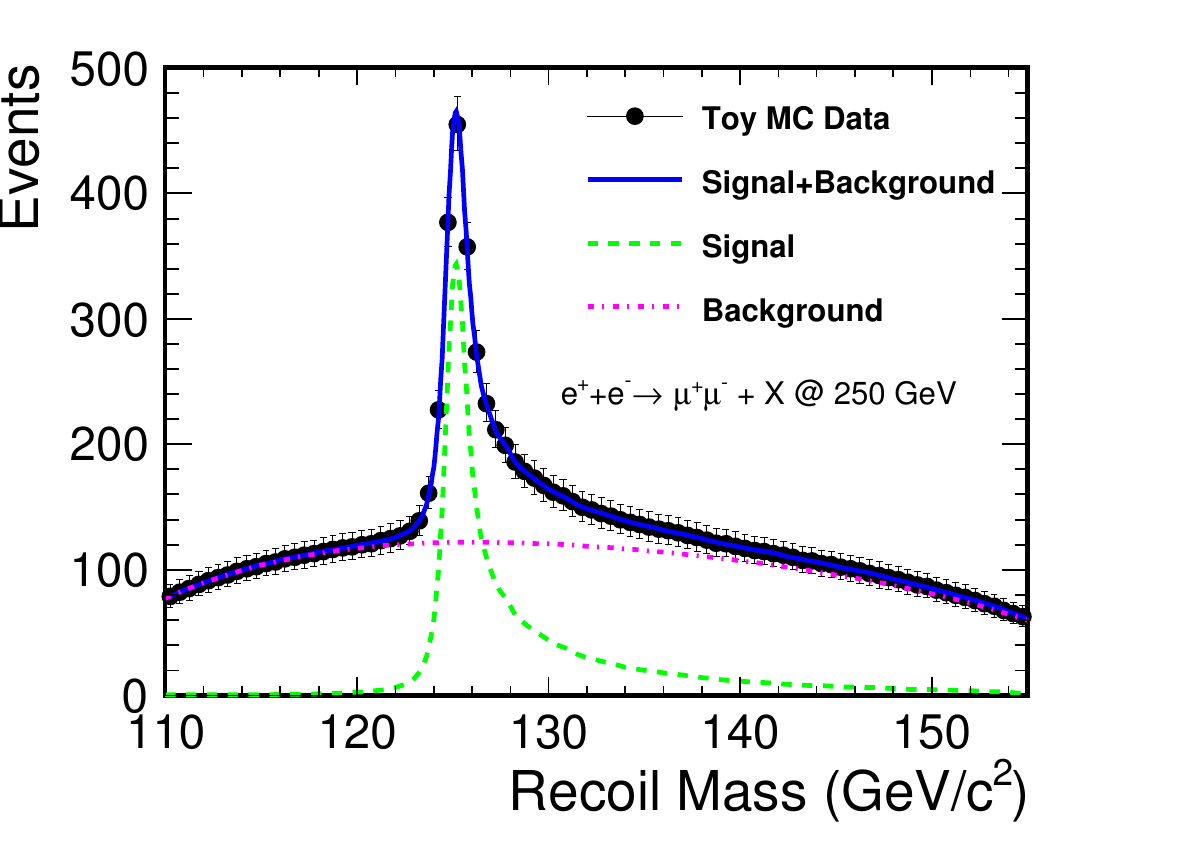}
      \end{center}
\caption{Determination of the Higgs boson mass at 250~GeV from
  measurement of the $Z$ recoil mass in $\ee\to ZH$ with $Z \to
  \mu^+\mu^-$, from \cite{Yan:2016xyx}.  The plot shown corresponds
  to an integrated luminosity
  of 250~fb$^{-1}$, to be compared to 3~ab$^{-1}$ for the LCF program
  at 250~GeV.}
\label{fig:Higgsrecoil}
\end{figure}

With this context, let's begin our discussion of the expectations for
the measurements of Higgs boson couplings -- in particular, the
couplings accessible in Higgs boson decay.  The dominant cross
sections for Higgs boson production at $\ee$ colliders are
Higgstrahlung, $\ee\to ZH$, and $WW$ fusion, $\ee \to \nu\bar\nu H$.
The cross sections for these reactions in the CM energy range of Higgs
factories are shown in Fig.~\ref{fig:Higgscx}.   The cross sections for
these two reactions cross over at about 470~GeV.

In this section, I will discuss the study of the Higgs from the
Higgstrahlung reaction, optimized at the peak of the cross section at
240-250~GeV.  This working point has important advantages for the
Higgs study.  At this point, each Higgs boson is tagged by a recoiling
$Z$ boson with a lab-frame energy of 110~GeV.  The peak in the $Z$
energy distribution, shown as a recoil mass of the Higgs boson, is
presented in Fig.~\ref{fig:Higgsrecoil}.   Almost every $Z$ boson in
the peak is recoiling against a Higgs boson observed in  some
possible decay mode.  The background under the peak comes mainly from
$\ee\to ZZ$ plus radiation; it is smooth and also precisely
calculable. Using only the observed $Z$ boson, it is possible to
measure the absolute Higgstrahlung cross section and, using the
best-defined $Z$'s with decay to $\mu^+\mu^-$, to measure the Higgs
mass to 10~MeV  ($10^{-4}$ precision).  Looking on the other side of
tagged events, it is possible measure ratios of branching ratios,
including branching ratios to invisible and exotic modes, in an almost
completely model-independent way.

The one missing piece of information from this study is the Higgs
total width. The predicted width in the SM is 4~MeV, too small to be
extracted from the recoil line shape.
If all exotic decays of the Higgs (including invisible
and partially-invisible modes) can be observed,  the total width can
be inferred from the absolute value of the Higgstrahlung cross section
and the constraint that the branching ratios sum to 1.   If one lacks
confidence that all ``unclassified'' Higgs decay modes can be
identified, it is still possible to determine the total width from a
fit to some theoretical model.

Often, this model is provided by the kappa framework, in which each SM
decay coupling $g(HAA)$ is multiplied by a factor $\kappa_A$.
However, this approach is model-dependent and also gives some
counterintuitive results.   In typical fits, the $HZZ$ coupling has a
tiny uncertainty while the $HWW$ coupling has an enhanced
uncertainty.  This defies our intuition, shaped by precision
electroweak measurements, that electroweak interactions are tightly
constrained by custodial symmetry that relates the couplings of $W$
and $Z$.   The asymmetry is caused by the assumption of the kappa
framework that the $HZZ$ coupling has the same Lorentz structure as
in the SM, while, in general,  additional structures can appear.

In \cite{LCSMEFT}, we suggested a better approach, based on Standard
Model Effective Field Theory (SMEFT)~\cite{SMEFTreview}.  SMEFT will
appear several times in our discussion, so let me give a brief
description of it here.  SMEFT writes
the true Lagrangian of nature as an expansion about the SM
Lagrangian.  The fields are taken to be those of the observed
particles of SM, with possible heavier particles integrated out.  The
interactions are the most general ones that obey the SM gauge
symmetries.  The series is ordered by operator dimension,
\beq
\L = \L_{SM} + \sum_i {c_i\over \Lambda^2} {\cal O}_i^{(6)} +
  \sum_j {d_j\over \Lambda^4} {\cal O}_j^{(8)} + \cdots \ .
    \eeq{SMEFTL}
Here $\L_{SM}$ is the SM Lagrangian, which is actually the most
general Lagrangian satisfying these criteria with only operators of
dimension 4 and below.   The additional terms listed are the
contributions from operators of dimension 6 and 8.  Operators of odd
dimension necessarily violate lepton or baryon number, so I have
omitted them for the discussion of Higgs properties.   The parameter
$\Lambda$ is the mass scale of new, Beyond-SM, particles.  If $\Lambda > 1$~TeV,
the dimension 6 operators will already be small corrections to the SM
and higher-dimension operators will be further suppressed.

A problem in working with SMEFT is that the total number of operators
appearing at each level is large.   For the SM with 3 generations,
already at dimension 6 there are 2499 operators.  However, for the
analysis of Higgs couplings, assuming that electroweak perturbative
corrections to the SMEFT corrections
can be ignored, we can restrict ourselves to operators that
appear in Higgs factory processes at the tree level, of which there
are only 18.  Adding two parameters for the Higgs invisible branching ratio
(measured) and the ``unclassified'' branching ratio (assumed
unmeasured) gives a closed fit.  Since \leqn{SMEFTL} is the full
Lagrangian of nature, we can add precision measurements from other
processes, including $\ee\to W^+W^-$ and precision electroweak
observables.   With this very relevant information, the fit
tightly constrains the Higgs boson
total width and the absolute values of all couplings.   This procedure
can make use of beam polarization, if it is available, to
resolve the coefficients of chiral
operators; the kappa fit, with its restrictive assumptions,
ignores the effect of beam polarization.  Unsurprisingly, the use of beam
polarization turns out to give a significant advantage.

\begin{table}[t]
     \begin{center}
  \begin{tabular}{lcccc}
 coupling   &   LCF 250 &  LCF 550   &  FCC-ee  &   CEPC \\ \hline
    $bb$    &   0.72 &  0.38 &  0.42 &  0.38 \\
    $cc$    &     1.45   & 0.87     &   0.78      &    0.64     \\
   $gg$    &      1.31         &    0.70     &    0.68      &    0.58        \\
   $\tau\tau$    &    0.83           &  0.53       &   0.48       &     0.42       \\
   $WW$    &      0.34         &    0.22     &   0.26       &    0.28        \\
   $ZZ$    &        0.34       &   0.22      &    0.26      &      0.27      \\
   $\gamma\gamma$    &   1.02            &   0.94      &     0.94
                                                & 0.91 \\
    $\mu\mu$    &   3.87            &  3.53      & 3.33       & 3.01
   \\ \hline
   $\Gamma_{tot}$    &     1.39          &   0.85     &     0.82     &    0.76        \\
   $BR$(invis.)    &        0.36       &    0.31     &     0.25     &    0.26        \\
   $BR$(other)    &     1.53          &    1.11     &    0.78      &   0.67        \\
 \end{tabular}
     \end{center}
     \caption{Expected 1 $\sigma$ precisions ($\Delta g/g$,
       in \%) for Higgs boson couplings from
    the expected run plans of linear and circular Higgs factories,
    from the SMEFT fit described in the text.   The last two lines are
    95\% CL upper limits on branching ratios.   The uncertainties in
    the $HWW$ and $HZZ$ couplings improve to about 0.15\% if all
    exotic decays of the Higgs boson can be observed and there is no
    ``unclassified'' category of decays.  The first two columns of
    this table were presented in \cite{LCVision}.}
\label{tab:Hcouplings}
\end{table}

\begin{table}
     \begin{center}
  \begin{tabular}{lccc}
 coupling   &   LCF 250 &  LCF 550   &  FCC-ee \\ \hline
    $bb$    &      0.59  &   0.27 &  0.30 \\
   $gg$    &      0.82   &   0.53 &    0.44  \\
   $\tau\tau$    &   0.69 &     0.45 &    0.38   \\
   $WW$    &      0.42   &   0.13 &  0.21   \\
   $ZZ$    &           0.50  &   0.13   &   0.21\\    
   $\gamma\gamma$    &   0.82 &     0.53 &  0.44\\
 \end{tabular}
     \end{center}
     \caption{Expected 1 $\sigma$ precisions ($\Delta g/g$,
       in \%) for Higgs boson couplings from
       the expected run plans of linear and circular Higgs factories,
       from the SMEFT fit presented in Figure~3.6 of
       \cite{deBlas:2025gyz}.  This fit includes some NLO SMEFT
       corrections.  It assumes that there are no invisible or exotic
       Higgs decays.}
\label{tab:HcouplingsBB}
\end{table}

It is straightforward to use this method to compare the expectations for
the precision in Higgs coupling measurement between linear and
circular colliders.  In Table~\ref{tab:Hcouplings}, I present the
expected 1 $\sigma$ precisions (in \%) for LCF, FCC-ee, and CEPC.
I assume the same detection efficiencies and background subtractions for
all colliders, based on the full simulation studies done for ILC.
This makes clearer that  the differences in the results that are due to the
different run plans, that is, from the integrated luminosities at the
various energy stages and from beam polarization for LCF.    For LCF,
I show the results for the initial stage at 250~GeV with 3~ab$^{-1}$
of data and for the full program of including 8~ab$^{-1}$ of data at
550~GeV.
For FCC-ee and CEPC, the main sources of data are the runs at 240~GeV,
but the estimates given include also anticipated running at the top
quark threshold and just above.  For FCC-ee, I use the run
plan in~\cite{FCCee2}, with 10.8~ab$^{-1}$ of data at 240~GeV.
For CEPC, I use the run plan 
in \cite{CEPCSnowmass}, with 2 detectors and a total of 20~ab$^{-1}$
of data at 240~GeV. All estimates include expected results from HL-LHC
on the $H\to \gamma\gamma$ and $H\to \mu^+\mu^-$ decays.
The final results for the expected precisions
are remarkably similar.   In all cases, the uncertainties are
dominated by statistical errors, with systematic effects well under
control.

For comparison, I present in 
Table~\ref{tab:HcouplingsBB}
the corresponding results from the
SMEFT fit carried out in the Briefing Book~\cite{deBlas:2025gyz}.
This fit follows a similar
strategy with some differences in detail. More operators are included,
but also more pieces of data constraining them.   Most importantly, it
assumes that the Higgs boson has no exotic decays.   This table shows
the same general trends 
and the essential equivalence of the full programs of the
linear and circular Higgs factories for these coupling measurements.

  \begin{figure}
    \begin{center}
      \includegraphics[width=0.48\linewidth]{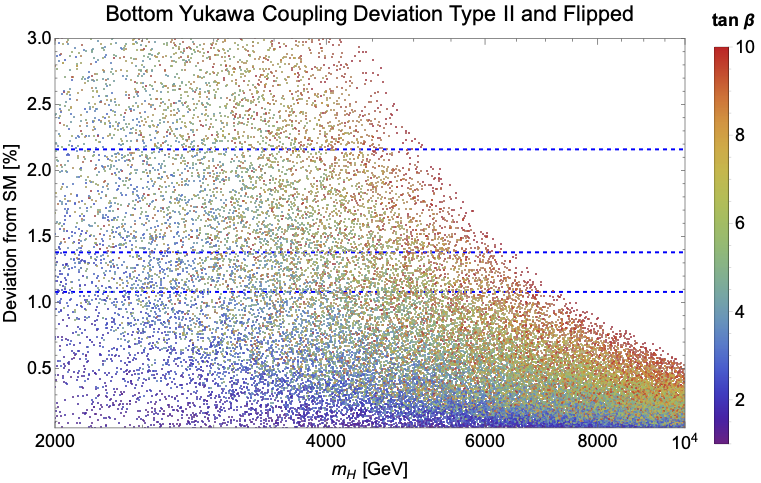}\
      \ 
     \includegraphics[width=0.48\linewidth]{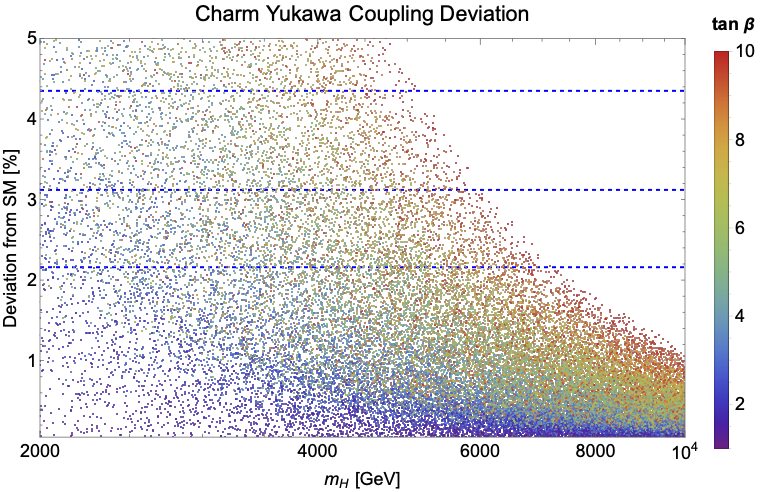}
      \end{center}
\caption{Scatter plots from a scan of the parameter space in Two Higgs
  Doublet (THDM) models, showing the relation between the heavy Higgs boson
  mass and the fractional deviation of a Yukawa coupling from its SM
  value.  Left: the deviation of the $b$ quark Yukawa coupling in
  the conventional Type II THDM; Right: the deviation of the $c$ quark
  Yukawa coupling in the flavorful THDM model of
  \cite{Altmannshofer}.   The horizontal dashed lines show
  3$\sigma$ deviation lines for stages of the LCF program.  The bottom
  such line shows the 3$\sigma$ sensitivity for the full Higgs factory
  program, either LCF
 or FCC-ee. From~\cite{DevinKamal1}.}
\label{fig:THDM}
\end{figure}

 \begin{figure}
    \begin{center}
      \includegraphics[width=0.48\linewidth]{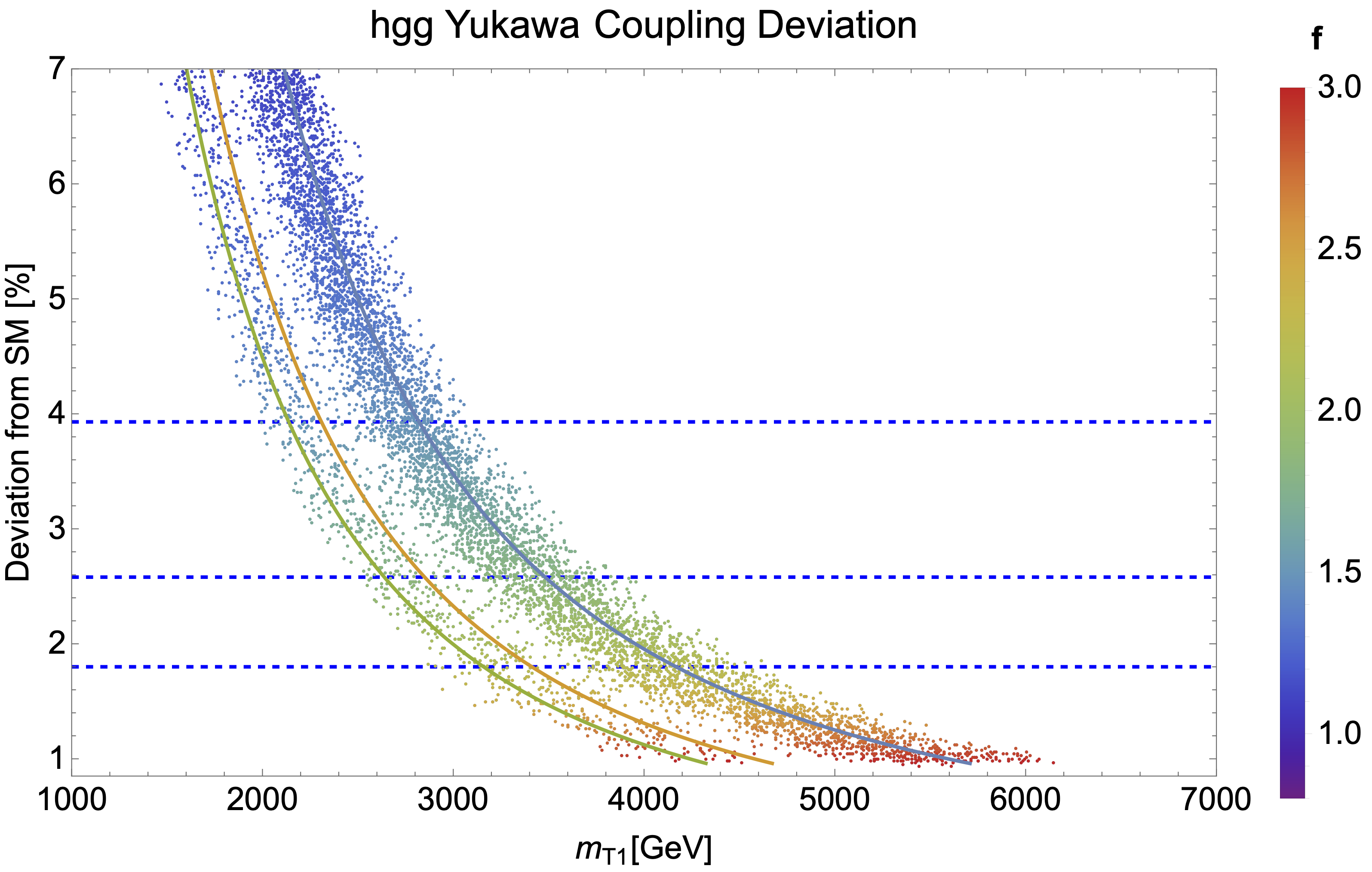}\
      \ 
     \includegraphics[width=0.48\linewidth]{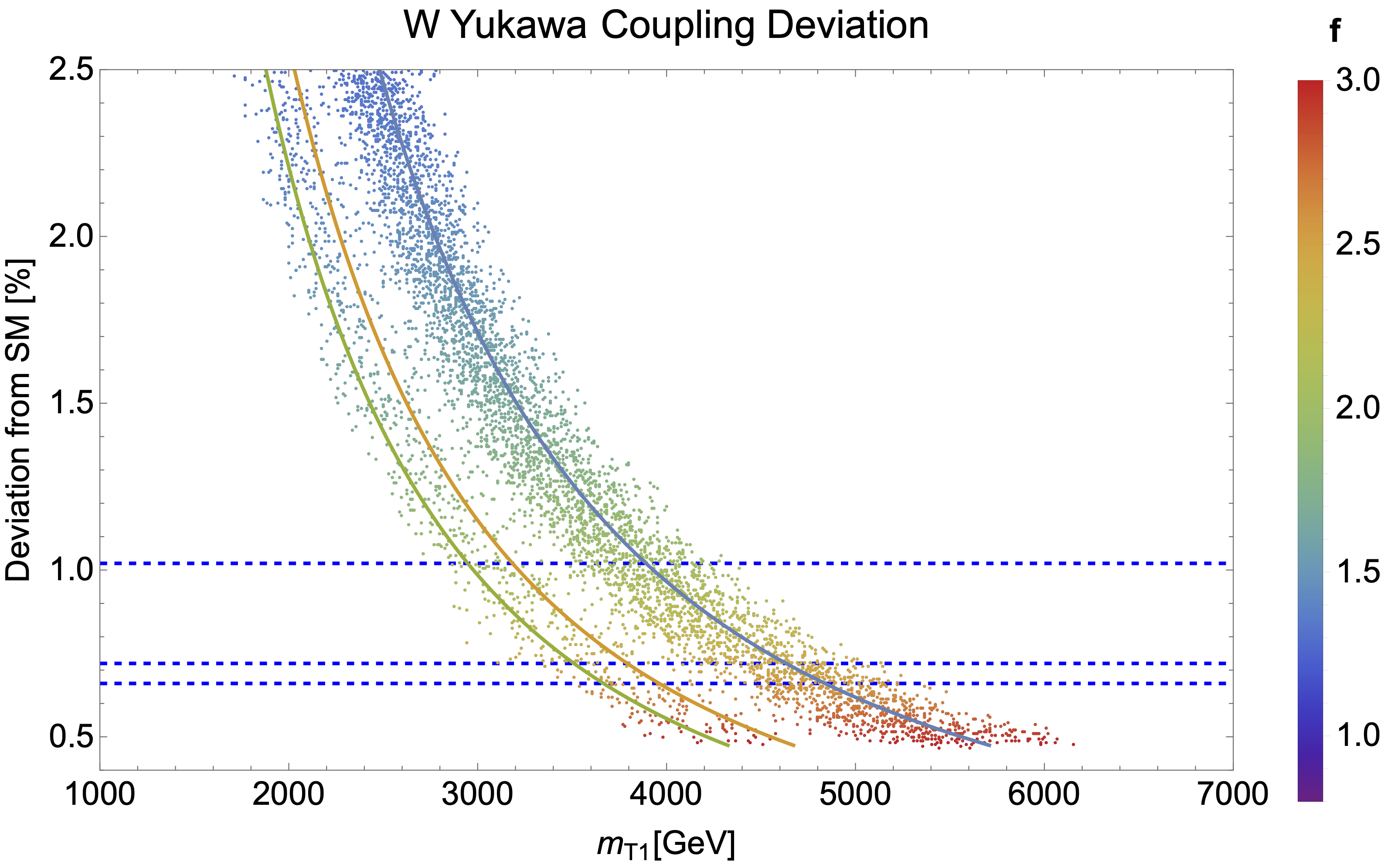}
      \end{center}\caption{Scatter plots from a scan of the parameter space in the
  Little Higgs model~\cite{LittleHiggs},
  showing the relation between the lightest vectorlike top quark
  partner mass
  and the 
  deviation of  Higgs  couplings from their SM
  values.  Left: the deviation of the gluon coupling to the Higgs
  boson; Right: the deviation of the $W$ boson coupling to the Higgs
  boson.  The solid lines show three different scenarios for the
  masses of heavier top quark partners.
   The horizontal dashed lines show
  3$\sigma$ deviation lines for stages of the LCF program.   The bottom
  such line shows the 3$\sigma$ sensitivity for the full Higgs factory
  program, either LCF
 or FCC-ee. From~\cite{DevinKamal2}.}
\label{fig:LH}
\end{figure}

A question often asked about Higgs coupling precision is:   Yes, there
are improvements over what is possible at LHC, but is it possible to
access the effects of new particles?   Given the success of the SM and
the fact that, with a Higgs boson mass of 125~GeV, the SM could be
valid up to the Planck scale, it is not possible to guarantee a
discovery, either in direct particle searches or in precision
studies.  However, there are real opportunities for discovery
available.   Some general classes of models giving sensitivity to new
particles at the levels of precision given in
Table~\ref{tab:Hcouplings} are reviewed in \cite{Agnostic}.

Recently,
Devin Walker and his group have explored this further by carrying out
complete parameter scans for some of these models.   In
\cite{DevinKamal1},  this group scanned the parameter space of
two-Higgs doublet models, giving particular attention to the ability
of these models to induce relatively large deviations from the SM
predictions for 
Higgs Yukawa couplings.  Figure~\ref{fig:THDM} shows two examples
  from this study, the possible deviations of the $b$ quark Yukawa
  coupling in standard Type II THDM modes, and the possible deviations of
  the $c$ quark Yukawa coupling in the ``flavorful 2HDM models'' of
  Altmannshofer and collaborators~\cite{Altmannshofer}.   The bottom
  dotted line indicates 3$\sigma$ sensitivity from Higgs precision in
  this variable.  (Typically, in these models, a fit to all observed
  deviations then  gives
  an overall significance of 5$\sigma$.)  The discovery reach for
  heavy Higgs bosons extends above 5~TeV in a large part of the parameter
  space.
  Figure~\ref{fig:LH} shows a
  complete parameter scan of the representative Little Higgs model of
  \cite{LittleHiggs}, showing deviations from the SM in the $Hgg$ and
  $HWW$ couplings~\cite{DevinKamal2}.  Here, the particles responsible for the deviations
  are heavy vectorlike top quarks.   The possible  reach in mass is
  above 3~TeV.   In both cases, these reaches are well above the
  expected limits that would be obtained direct resonance searches at
  the HL-LHC.   There is
  opportunity for discovery here, and we should grasp at it.

\begin{figure}
    \begin{center}
      \includegraphics[width =0.9\linewidth]{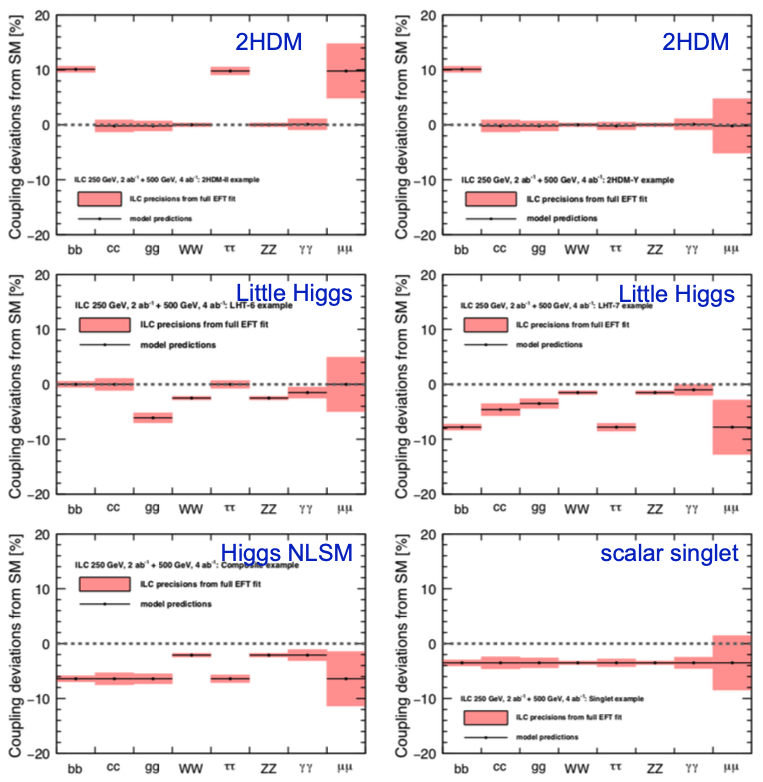}
      \end{center}
\caption{Relative deviations of the Higgs boson couplings in six
  diverse models of new physics. First row: 2-Higgs doublet models;
  Second row: Little Higgs models; Third row: a Composite Higgs model
  and a Scalar Singlet model. Error intervals shown are those for the
  ILC500. From \cite{ILC}; see \cite{LCSMEFT} for details on the
  specific models chosen.}
\label{fig:Higgspattern}
\end{figure}

  Surveying models of BSM physics more broadly, more such
  opportunities
  are available.  From model to model, the deviations occur in
  different Higgs boson coupling.   Many different Higgs couplings
  will be measured with high precision, and  so the pattern of deviations gives
  information about the higher energy theory that gave rise to them.
  Figure~\ref{fig:Higgspattern}, from \cite{ILC}, illustrates this.
  The precision study of Higgs couplings gives us the opportunity not
  only to discover deviations from the SM but also to gain important
  clues as to its origin.

  \section{Study of the $Z$ boson at high statistics}

The high luminosities available at lower energies at circular
colliders make it possible to
imagine a very high statistics program of measurements at the $Z$
pole, gathering about $5\times 10^{12}$ $Z$ decays (Tera-Z).   Even with the
lower luminosities available at linear colliders, it is possible to
gather a sample of $5\times 10^9$ $Z$ decays, an improvement of more
than 3 orders of magnitude over the LEP program, and with polarized
beams (Giga-Z).  It is interesting to discuss the potential for the
discovery of BSM effects with these large data set.

One goal of this program repeats that of LEP:   Test the SM by
measuring $\sin^2\theta_w$ to high precision.  In both programs, this
can be measured from electron left-right asymmetry of the $Z$ coupling,
$A_e$.   In the Tera-Z program this is done using the production angle
asymmetry of $Z\to \tau^+\tau^-$ decays; in the Giga-Z program, this
is done using the asymmetry of the total hadronic cross sections using
polarized beams.  Both techniques reach uncertainties $\Delta
\sin^2\theta_w/\sin^2\theta_w \sim 1.5\times 10^{-5}$, an improvement of
more than an order of magnitude from the current uncertainty.   In the
Tera-Z program, one can also access $A_\ell$ from the forward-backward
asymmetries in $\mu^+\mu^-$ and $b\bar b$ production.  I will comment
on these below.

\begin{figure}
    \begin{center}
      \includegraphics[width=0.3\linewidth]{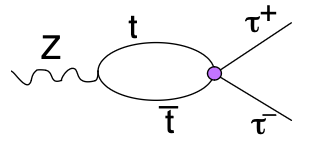}
      \end{center}
\caption{A representative diagram that includes a dimension-6 SMEFT
  operator and modifies the $Z$ boson coupling to fermions at 1-loop
  order in electroweak interactions.}
\label{fig:ZSMEFT}
\end{figure}

However, this program also contains a  more powerful probe of
Beyond-SM physics.   The precisions are such that the measurements are
sensitive to loop corrections from new heavy particles that are also
subject to SMEFT suppressions $v^2/M^2 \sim
1\%$\cite{Celada,Allwicher,Maura}.
A typical diagram
including a dimension-6 SMEFT operator in a $Z$ radiative correction
is shown in Fig.~\ref{fig:ZSMEFT}.   For $M \sim 1$~TeV, this would be
a deviation from the SM precision electroweak predictions of order
$10^{-4}$.   With the very high statistics available from a $Z$ pole
program, these effects give sensitivity to new physics beyond the SM
at the 10~TeV scale and beyond.    The potential sensitivity is
comparable to that available from measurements of heavy quark weak
interactions.  But these effects do not rely on flavor violation or
violation of weak interaction universality.   They are produced by
generic modifications at the BSM energy scale.

Still, there are two qualifications that must be given to the promise
of the previous paragraph.   First, though new physics effects can
come from a very wide range of operators, the strongest observable
effects occur in a relatively small number of electroweak
observables.   The most important ones are:    (1) the total width 
$\Gamma_Z$,  (2) the hadron to lepton ratios  $R_e$, $R_\mu$,
$R_\tau$, (3) the ratio $R_b$ of $b$ decays to all hadronic decays, which
specifically probes operators of the 3rd generation, and (4) $m_W$ or
the $T$ parameter, which probe custodial symmetry violation and 
might also signal effects of the heavy top quark.  In all cases,
the effects would be tiny modifications of these quantities, which
must be compared to high-precision predictions of the SM values.

This raises the second qualification:   Can we produce the SM
predictions with sufficient precision, and defend these in the case
that an anomaly is observed? This is the question of the
understanding and reliability of systematic uncertainties.  The issues
of raised by this question were clarified in the studies of Ayres Freitas for
the 2026 ESPP update~\cite{Freitas} and are discussed
in Section 3.4 of the Briefing Book~\cite{deBlas:2025gyz}.
Freitas distinguished two sources of systematic error:  first, the
uncertainties from omitting terms beyond a given order in electroweak
perturbation theory, second, the uncertainties in the modeling of
physics events and the conversion of what is actually measured to
ideal ``pseudo-observables'' to be compared to the results of
calculation.   I will give my own opinion on these sources, which differs
from the attitude taken in the Briefing Book.

On the first issue, the truncation of perturbation theory, I have
great confidence that theorists will achieve sufficiently accurate
calculations.  Essentially, precision electroweak observables and the
electron structure functions must be
computed to N$^3$LO order with general nonzero masses.   This is a
major challenge.  However, theorists have met a
similarly difficult
challenge in improving methods for perturbative QCD to provide N$^3$LO
and multijet cross sections needed for LHC.  Also, this is a well-defined
problem, so that, once it is solved with new insights, many groups will
verify the results.

On the second issue, I have much less confidence.  The sources of
uncertainty here come from  non-perturbative effects as they are
modeled in the widely used Monte Carlo programs.
This is not a matter of lacking statistics; what we lack is an
accurate theoretical understanding of the non-perturbative stage of
hadronization.  It is noteworthy, and somewhat depressing, that we still
estimate systematic errors from hadronization from the differences
between the results of PYTHIA and HERWIG.
Today, there is no strategy  improve this situation.
We should treat uncertainties from this source more conservatively.

The authors of the Briefing Book estimated ``conservative''
and ``aggressive''
systematic uncertainties on the important precision electroweak
measurements.   The ``conservative'' estimates
represent improvements, though not revolutionary
ones, in the current state of
the art.  The ``aggressive''
uncertainties are, essentially, what is needed to overcome the
barriers described in the previous paragraph.
They include the improvement of uncertainties in
nonperturbative hadronization by a factor 50 and setting to zero
uncertainties that we currently do not know how to estimate.  This
seems to overlook the requirement of defending any claim of violation
of the SM against the skepticism of our colleagues.

The ``conservative''  estimates of theory uncertainties are listed
in Table~\ref{tab:EWPOsyst}, along
with the statistical and experimental systematic uncertainties
expected from the $Z$ pole programs of LCF and FCC-ee.  It is
important to note that these theory systematic uncertainties are large
compared to the claimed experimental systematics, especially for the
Tera-Z program.  The resulting uncertainties on the $Z$ couplings to
fermions are given in Table~\ref{tab:Zcouplings}.  In contrast, 
the projections given in the Briefing Book for the uncertainties in Higgs
couplings (Table~\ref{tab:HcouplingsBB}),  which are not at this very
high level of precision,  are quite robust to the
  choice  of ``conservative'' or ``aggressive'' uncertainties. The
  differences are at the 20\% level in the estimated uncertainty on
  the Higgs coupling for the $Z$ and $W$
and at the percent level for the other cases.  This same
  robustness appears for the results on top quark interactions
  discussed in the following section.

\begin{table}
     \begin{center}
  \begin{tabular}{lccc}
    observable  &  Tera-Z  &  Th. Systematics   & Giga-Z  \\ \hline
$Z$ width: \\
   \ \  $\Gamma_Z$   &   0.05  & 0.2 [2.2] & 5 \\ \hline
    Ratios of BRs:  \\
\ \ $R_e, R_\mu$, $R_\tau$   &  0.3      & 1.2  [3.2]    &    13   \\
 \ \    $R_b, R_c$     &    0.20/0.26        &  20/100 &    9/29  \\ \hline
    $\sin^2\theta_w$:\\
 \ \  $ A_e(A_{LR}) $ &          &   19       &     19       \\
  \ \  $ A_e(\tau\ \mbox{pol}) $ &   14       &  27     &        \\
 \ \ $ A_{FB,b},  A_{FB,c} $   &      5/9    &   31/31  &        \\
\ \ $ A_{FB,\mu},  A_{FB,\tau} $   &     20      &  230     &     \\
    \hline
    $W$ mass:           \\
 \ \   $m_W$         &  0.3   &  0.7       &   3 \\
  \end{tabular}
     \end{center}
     \caption{Expected systematic uncertainties for the major
       precision electroweak observables.  Values for FCC-ee are taken
       from
       \cite{BlondelSelvaggi}; values from LCF are taken from
       \cite{LCVision}.
       Theory systematics are taken from~\cite{Freitas}.  All uncertainties are given
       as relative uncertainties in units of $10^{-5}$. Values in
       brackets are the uncertainties generated by the uncertainty in
       the measured value of
       $\alpha_s$, assumed to be $\Delta \alpha_s = 10^{-4}$. }
\label{tab:EWPOsyst}
\end{table}

\begin{table}
     \begin{center}
  \begin{tabular}{lcc}
    $Z$ coupling to:  &  LCF 250  &  FCC-ee 240  \\ \hline
$e_L$           &      0.015   &      0.011   \\
$e_R$            &             0.021   &     0.015   \\
$\tau_L $        &            0.022    &      0.013   \\
$\tau_R$         &            0.030   &     0.019    \\
$c_L $              &         0.054   &     0.051   \\
$c_R $            &           0.082    &      0.071  \\
$s_L $             &           0.043   &       0.040   \\
$s_R$              &           0.54    &      0.34   \\
$b_L$              &           0.023   &      0.019    \\
$b_R$             &           0.26    &       0.33   \\
   \end{tabular}
     \end{center}
     \caption{Expected uncertainties in the $Z$ couplings to  fermions
       from the LCF 250 and FCC-ee 240 experimental programs from the
       SMEFT fit in the Briefing Book~\cite{deBlas:2025gyz}, under the
      ``conservative'' assumptions for theory
       systematic uncertainties.  In both programs, the strongest constraints on
       these couplings come from  the $Z$ pole measurements.}
\label{tab:Zcouplings}
\end{table}

Even with the ``conservative''
theory systematic uncertainties, the power of the $Z$ pole measurements
is very impressive.  This is often presented as a table or graph of
sensitivity scales $\Lambda$ for higher-dimension SMEFT operators.
In \leqn{SMEFTL}, the coefficient of each
dimension-6 operator is written as $c_i/\Lambda^2$, where $\Lambda^2$
is a mass scale.  In an underlying BSM model, effects of new heavy
particles generate corrections to the SM, and these are described by
specific higher-dimension operators included in
the SMEFT Lagrangian.   In a fit to data with a Lagrangian  including some
number of these operators, the standard  error $\sigma$ on each coefficient
gives an idea of the masses of the new particles generating this
operator
to which that data set is
sensitive.   That scale would be roughly $\Lambda \sim 1/\sqrt{\sigma}$ for
operators generated by new strong interactions and
$\Lambda/\sqrt{\alpha_w}  \sim
\sqrt{\sigma}$, for operators generated by electroweak radiative
corrections,
giving $\Lambda$ values a factor 6--10 smaller.  

This can be viewed in two ways.  First, one can include
all dimension-6 operators that respect basic symmetries of the
theory.  The fit in the Briefing Book assumes SM gauge
symmetry, CP conservation, and a $U(2)$ flavor symmetry for the
lighter two fermion generations.  This leads to a Lagrangian with 124 
dimension-6 operators. Of these, the 84 operators with unsuppressed
couplings to precision observables at the tree or 1-loop level were
included in the analysis.
For each operator, the uncertainty in the operator coefficient given
by the global if can be converted to a mass scale using the relations 
just described.   This estimates the mass scale for which an observed
deviation in the data can be shown to be induced by that operator.
From the electroweak observables, this fit gives
sensitivities to new particle masses in the few-TeV region.
I will give some examples below in
Table~\ref{tab:topoperators} in the following section.

However, one
can also probe the data in another way,
by including only one dimension-6 operator in
the fit.   This would represent a particular mode of violation of the SM.
In that case, the uncertainty in the SMEFT coefficient for that
operator  and the
associated $\Lambda$ scale estimates
the sensitivity to new physics from all
possible operators creating similar effects.  That is, these
single-operator fits estimate  in more general way the mass scale for
to which the data is sensitive.
In that case, the fits to expected
precision electroweak data give sensitivity to scales $\Lambda\sim
20-60$~TeV.  Specific values for different top quark operators are
also given in Table~\ref{tab:topoperators}.

Deviations from the SM predictions for precision electroweak
observables could well be the first evidence the that the SM must
be corrected by new physics at a high mass scale.   This is a
capability within the reach of both circular and linear Higgs factories.

\section{Higgs measurements above the top quark threshold}

Finally, I will discuss the Higgs factory program at energies above the 
top quark threshold.    This region is often omitted from discussions
of the physics of Higgs factories, but in fact it allows additional
important measurements whose results are essential to the Higgs
story.

First, running at high energy will allow us to redo the measurements
of the Higgs couplings listed in Table~\ref{tab:Hcouplings}.  
As I pointed out above, the $WW$ fusion reaction becomes the dominant
source of Higgs boson production above 400~GeV, so the new Higgs sample
will be dominated by this new reaction, with different experimental
issues and sources of uncertainty. This will be an important check on
any anomalies discovered in the program at the Higgs threshold.
Assuming compatibility of the two sets of results, the improvement in
the uncertainties on Higgs couplings has already been shown in
Table~\ref{tab:Hcouplings}.

Second, the top quark Yukawa coupling can be measured at CM energies
of 550~GeV and above in the reaction $\ee\to t\bar t H$.  From ILC and
CLIC studies, we expect an uncertainty of 2.8\% with 8
ab$^{-1}$ of data at 550~GeV and 1\% at with 8 ab$^{-1}$ of data at
1~TeV~\cite{LCVision}.   This latter value matches the statistical
uncertainty on the top quark Yukawa coupling eventually
expected from the FCC-hh ~\cite{FCCee2},
but in an environment in which the systematic errors are subdominant.

  \begin{figure}
    \begin{center}
      \includegraphics[width=0.7\linewidth]{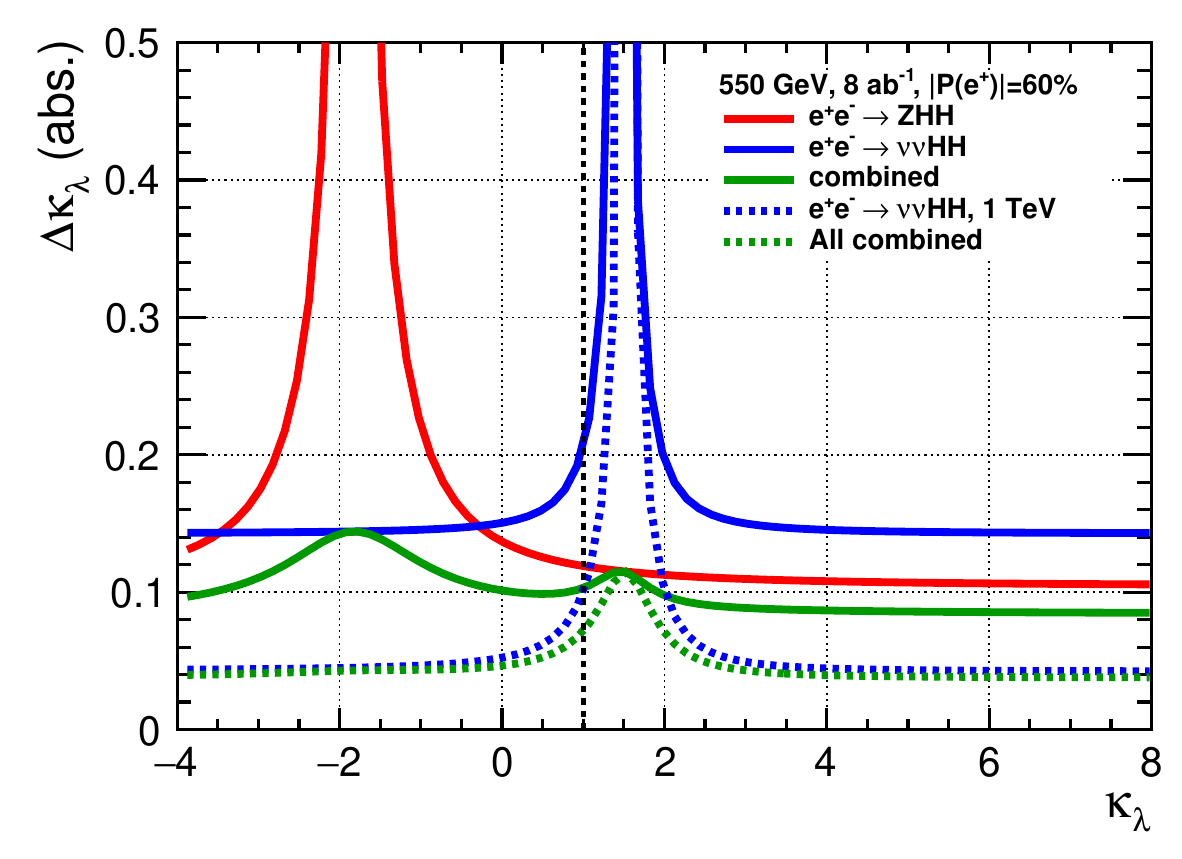}
      \end{center}
 \caption{ Absolute precision on the deviation of the Higgs
  self-coupling from its SM value ($\Delta \kappa_\lambda$) expected
  from the measurement of Higgs pair production in the 
  ZHH and WW-fusion processes at  550 GeV, showing the
  complementary information from the two production modes, from \cite{LCVision}.}
\label{fig:kappalambda}
\end{figure}

Third, a Higgs factory operating at 550~GeV and above can measure
Higgs pair production, which directly accesses the Higgs boson
self-coupling.  An $\ee$ collider at 550~GeV can
measure Higgs pair production in two reactions, $\ee\to ZHH$ and the
$WW$ fusion reaction $\ee\to \nu\bar\nu HH$.  With 8~ab$^{-1}$, the
expected precision in the measurement of the Higgs self-coupling is
11\%.  This is illustrated in Fig.~\ref{fig:kappalambda}~\cite{LCVision}.   However,
this is not the most important conclusion from this figure.  Notice
that the combined uncertainty in
the deviation from the SM ($\Delta\kappa_\lambda$) is
uniform as a function of the true
underlying value of the self-coupling.   The reason for this is that
the amplitude for Higgs pair production is a sum of the contribution
from the self-coupling and contributions due to other SM processes.
In the $ZHH$ reaction, these amplitudes are in constructive
interference; in the $WW$ fusion reaction, they are in destructive
interference.  For a positive deviation of $\kappa_\lambda$, we should
observe an increased value of the $ZHH$ cross section and a decreased
value of the 
$WW$ fusion cross section in the same experiment.   This
observation is possible
only at an $\ee$ collider.  Observing both effects might be necessary
to provide a
credible discovery that the Higgs  self-coupling differs from the SM
prediction.   The figure also shows the improvement in precision (to
about 5\% over most of the region) available with an additional 8
ab$^{-1}$ at a CM energy of 1~TeV.

\begin{figure}
    \begin{center}
      \includegraphics[width=0.7\linewidth]{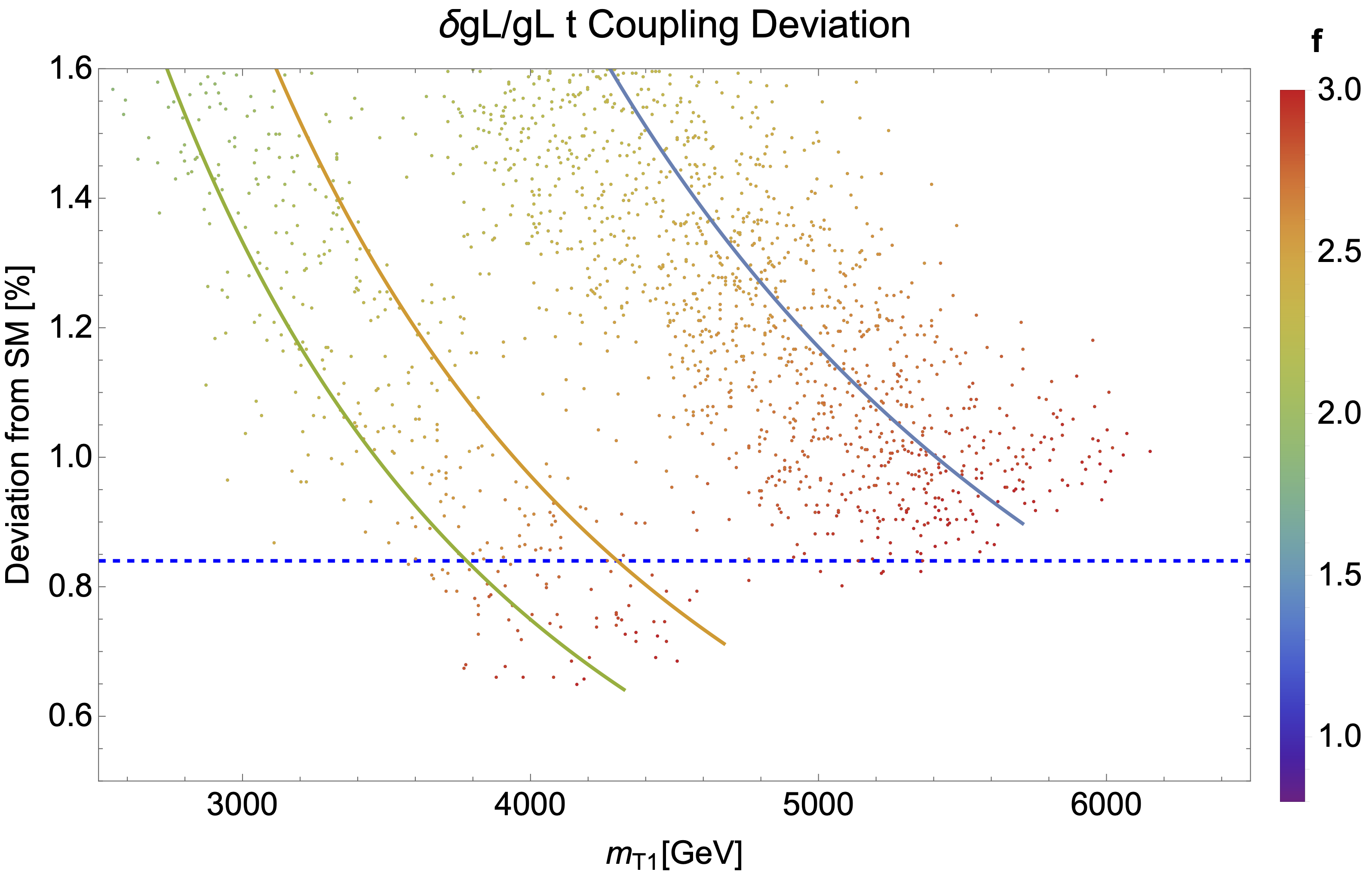}
      \end{center}
 \caption{Scatter plot from a scan of the parameter space in the
  Little Higgs model~\cite{LittleHiggs},
  showing the relation between the lightest vectorlike top quark
  partner mass and the deviation from the SM value of the $Z$ coupling
  to the $t_L$.  (There is no effect on the $t_R$ in this model.)
  The solid lines show three different scenarios for the
  masses of heavier top quark partners.
   The horizontal dashed line shows the 
  3$\sigma$ sensitivity for the LCF 550 program. From~\cite{DevinKamal2}.}
\label{fig:topEW}
\end{figure}

Finally, running of an $\ee$ collider well above the top quark
threshold allows the precision determination of  the electroweak
couplings and possible additional interactions of the top quark.
In composite models of the Higgs boson, it is typical that the top
quark must also couple with some strength to the compositeness
interactions in order to have its large Yukawa coupling.  This affects
the $W$ and $Z$   couplings of the top quark at the several percent level
for top partners near the current LHC limits. The modification is
directly visible in the production cross section for top quark pairs,
which goes through $s$-channel $\gamma$ and $Z$ exchange.   This effect is
relevant in the Little Higgs model studied in Fig.~\ref{fig:LH}.  The
scatter plot for the deviation of the $t_L$ coupling to the $Z$
over the parameter space of
the model is shown in Fig.~\ref{fig:topEW}~\cite{DevinKamal2}.  The horizontal dotted
line shows the  3$\sigma$ sensitivity expected for the LCF program at
550~GeV\cite{LCVision}.   Modifications of the top quark $Z$
couplings are generically present at this level in Little Higgs models
and also in 5-dimensional Randall-Sundrum
models of a composite Higgs.   Many more examples are provided in~\cite{LCVision}.

More generally, the CM energy dependence , angular dependence,
and beam-polarization
dependence of the reaction $\ee\to t\bar t$ can all be sensitive to
modifications induced by  dimension-6 operators.  With enough
observables measured, it is possible to distinguish the effects of
different operators.    This has been tested in full-simulation
studies carried out by the ILC and CLIC groups for event samples at
380~GeV, 550~GeV, and 1~TeV.   These studies are described in
\cite{Cornet-Gomez:2025jot,Durieux:2022cvf}, with some updated results
presented in \cite{LCVision}.   The method used is to fit to
simulation data the predictions from
a SMEFT Lagrangian with 21 dimension-6 operators
including the 11 most important operators bilinear in top quark
fields.  The covariance matrix obtained for the dimension-6 operator
coefficients was then used as an input to the larger
84-operator global fit carried out for the Briefing Book.
The
resulting uncertainties for the 11 top quark operators are given in
Table~\ref{tab:topoperators}.

\begin{table}[t]
     \begin{center}
  \begin{tabular}{lccccc}
operator   &   LCF 250 &  FCC-ee 240 & FCC-ee & LCF 550   &  LCF1000  \\ \hline
 \multicolumn{2}{l}{electroweak couplings:} \\\hline
 \ \ $C_{\phi Q}^-$  & 0.65 / 27.5  &  0.84  / 30.8     &  1.12 /  34.4      &
                                                                      1.37/ 29.7   &  2.48 / 34.1 \\
 \ \ $C_{\phi Q}^3 $ & 1.02 / 37.9  &  1.27  / 41.3   &   1.66 /  44.2
                                               &  2.94 / 39.0   &    4.43 / 41.9\\ 
 \ \ $C_{\phi t} $   &   0.86 / 18.5  &   0.95 /  22.4  &   1.46 /  27.8
                                               &  1.44 / 23.5    &     2.49 / 30.2\\ \hline
   \multicolumn{2}{l}{Yukawa coupling:} \\ \hline
    \ \ $C_{t\phi} $  & 0.70 / 2.63  &  0.76 /  2.92 &  0.85 /  2.99
                                               &  1.20 / 2.91 &  1.44 / 3.07\\  \hline
   \multicolumn{2}{l}{4-fermion operators:} \\ \hline
 \ \ $C_{Q\ell}^- $  &  0.98 / 30.5  & 1.21 / 37.8  &  4.01 / 45.8  &
                        7.49 / 65.2 & 15.5 / 95.2 \\
 \ \ $C_{Q\ell}^3 $  & 1.36 / 42.9   &  1.67 / 50.6         &  3.41 / 62.3
                                               &  8.93 / 85.1  & 18.1 /  127.     \\         
 \ \ $C_{te} $   &   0.78 / 14.1   &   1.03 /  24.4      &  4.07  /    26.2
                                               &  4.10 /  45.3   &      15.1 / 61.5   \\
    \ \ $C_{t\ell} $   & 1.28 / 14.3  &  1.38 / 24.8       &  5.29 / 27.7
                                        &   5.90 / 42.6    &     20.3 / 56.6\\
    \ \ $C_{Qe} $   &   2.88 /   32.3  &  3.00 /  28.8  & 3.63 / 31.4
                     &    10.9 / 62.4    &    18.3 /   85.0\\ \hline
 \multicolumn{2}{l}{magnetic moments:} \\ \hline
    \ \ $C_{tW} $  &   1.25 / 26.0  &  1.49 / 33.8 &  4.68 / 41.3
              &     5.75   /40.8  &     6.67 /44.4\\
    \ \ $C_{tZ} $   &  1.16 / 23.2  &  1.20 / 29.7 &   3.99 / 36.1
            & 7.42 /37.6 &      8.69 / 40.7\\
 \end{tabular}
     \end{center}
     \caption{New physics scales $\Lambda$ (in TeV) to which SMEFT fits are
     sensitive at successive stages of the LCF  and FCC-ee
     experimental programs.   The column ``FCC-ee'' includes the full
     FCC-ee program including the run at 365~GeV.   The column $LCF
     1000$ include a run with 8~ab$^{-1}$ at 1~TeV.
     These scales are given by the 1$\sigma$ uncertainties in SMEFT
     operator coefficients, as explained at the end of Section 4. The
     values in this table assume the ``conservative''assumptions for
     theory systematic uncertainties. In
     each pair, the first value is that given by the global fit and
     the second value is that given by a fit with only this
     dimension-6 operator.  The important top quark operators are
     classified by the corrections they give to the top properties in
     the SM:  modification
     of the electroweak couplings, modification of the Yukawa
     coupling, introduction of 4-fermion contact interactions,
     modification of the electroweak magnetic moments.  From \cite{deBlas:2025gyz}.}
\label{tab:topoperators}
\end{table}

This Table requires some explanation.  The operators  included in the
left-hand column are:  the 3 operators that shift the left- and
right-handed top quark electroweak couplings, the operator that shifts
the top quark Yukawa coupling, the 5 operators that provide 4-fermion
$e^+e^-t\bar t$ contact interactions, and the 2 operators that shift
the top quark $W$ and $Z$ magnetic moments.  All of these operators
can be induced by BSM top quark physics.  I have discussed the shift in the
electroweak couplings above.  The 4-fermion operators can be induced
by constituent exchange in composite top quark models, or by the
Kaluza-Klein vector bosons in a higher-dimensional model of  Higgs compositeness.
Four-fermion $e^+e^-t\bar t$ operators can also be induced as NLO
corrections to order-1 4-top contact interactions.   

The values given in the table  are the
uncertainties in the operator coefficients interpreted, as I explained
at the end of Section 4, as sensitivities to large values of the new
physics scales $\Lambda$ in the SMEFT Lagrangian.   The first number
of each pair is the sensitivity of this particular operator from the
84-operator fit.  The second number is the sensitivity from a fit that
includes only this operator.   This tests the ability of this operator
perturbation to signal a deviation from the SM.    To the extent that
these numbers are of comparable size, the fit indicates that the data
is able to pick out that particular operator as the one responsible for an
observed deviation.

The single-operator sensitivities in the first two columns are already
very impressive.   These are mainly obtained from the $Z$ pole
programs of the LCF and FCC-ee.
As I noted in Section 4, the Tera-Z program  gives higher sensitivities, but
only to the extent that the uncertainties presented
Table~\ref{tab:Zcouplings} are smaller. The next two columns include
the first stages beyond the top quark threshold.   The
sensitivities from direct $t\bar t$ observations surpass those from the
$Z$
pole measurements, and more substantially as the CM energy is raised.
The specificity  of the operator identifications is
still low, because, with only one CM energy,  there is a degeneracy
in the effects from the  top quark electroweak modifications and the top
4-fermion couplings. However, the electroweak modifications to the
cross section stay at a constant size as a function of CM energy,
while the effects of 4-fermion operators increase as $E_{CM}^2$.
Then, adding a data set at 1~TeV clearly distinguishes the two
effects, giving marked increases in the ability of the fit to pick out
the specific top quark operator responsible.    Also, to the extent
that the new physics induces 4-fermion interactions, the 1~TeV data
substantially improves the overall sensitivity to BSM physics.

This study addresses a number of goals.  It is important
to recognize that the SM is violated by BSM effects, more important to
demonstrate that this effect is specifically associated with the top quark,
and still more valuable to distinguish between the electroweak
couplings and 4-fermion   effects, which are characteristic of
different classes of models.  Table~\ref{tab:topoperators} shows that
the program of high-energy running of an $\ee$ collider will address all
of these questions.

At the CM energies where top quark observations dominate the SMEFT fit
results, those results again become relatively insensitive to theory
systematics, giving very similar results from the ``conservative'' and
``aggressive'' assumptions discussed in Section 4. 
I commented there that observations of tiny deviations in
the precision electroweak observables may not have sufficient
power to convince skeptics that the SM is in fact violated.   For BSM
physics associated with the third generation, observation of
deviations from the SM predictions of the top quark interactions would
strongly confirm any effect suggested by the precision electroweak
program.   This statement generalizes to other source of BSM physics,
since
high energy $\ee$ experiments will look for deviations in the complete
set of final states available through 4-fermion contact interactions.

This completes our survey of the physics opportunities of Higgs
factories. These include the ability to discover effects from 
extended Higgs bosons, top quark partners, and other BSM particle well
beyond the direct discovery limits of the HL-LHC, and the ability to probe
for new BSM interactions at scales of 40~TeV and above.  For lower energies, circular
colliders have an  advantage due to their higher luminosities, but
this advantage is surprisingly modest.
This is quantified above in
Tables~\ref{tab:Hcouplings}--\ref{tab:Zcouplings}.

The measurements described in this section on the top quark Yukawa
coupling, the Higgs self-coupling, and the search for compositeness in
top quark interactions add important
capabilities that are available at CM energies of 550~GeV and above, a
region that can be probed only with linear collider Higgs factories.
Probably the most important feature of measurements in this energy region is
that they can provide reactions distinct from those at lower energy
that are  sensitive to the
same models of BSM
physics, giving the 
ability to make multiple independent measurements of suggested new
physics effects.  Without these, it will be difficult to fulfill the high burden
of proof for a discovery based on precision measurement.   The FCC-ee
at CERN has powerful capabilities for discovery. But we may well need
this higher-energy program also to build a truly persuasive case for
new physics beyond the SM.

\section{The challenge for now}

The  construction of a Higgs factory to begin operation just after the
completion of the HL-LHC program is important for our
understanding of physics, but it is even more important for  the early
career scientists in our community.    The HL-LHC, planned to operate
in 2030, offers exciting opportunities for the discovery of new particles,
especially for color-neutral states such as extended Higgs states and
electroweakinos with
masses in the range of 1~TeV.   But we can already see the limits of
this program and the need to look beyond it.  A new major collider
takes 10 years to construct.  This will be the dominant machine at the
peak of your career.  It is time to start thinking about it now.

This is going to be fun.   The environment of an $\ee$ collider
is very different from that of the LHC.   There is minimal radiation
load on the detectors, no underlying event, no pileup.   The $\ee$
annihilation  events are simpler than those at the LHC.   Not only are the
particle multiplicities lower, but there is no forward fragmentation
lost down the beam pipe.  This is an environment in which it is
possible to make vertex, tracking, and calorimetry measurements that
are close to perfect.  But -- this point is very important -- this
does not  make experimentation easier, because great improvements in
event reconstruction are need to satisfy the goals of high precision
measurements.

It is likely that the Higgs factories will provide the only chance
that you will have in  career to design a new collider detector from
the ground up.  This detector should incorporate new technologies and
should be
adapted for  heavy use of  Machine Learning and AI.  This is a
fascinating challenge.  It is also one for which you are much better
prepared than your professors.

As a first step, you should study the work that has already been done
to design detectors specifically adapted to make high precision
measurements at $\ee$ colliders.  I recommend, in particular, close
study of the ILD Interim Design Report~\cite{ILDConceptGroup:2020sfq},
which gives a thoughtful approach to precision measurement with
currently available technologies. Detectors meant to operate at Higgs
factories in the 2040's should go even further to incorporate new
ideas.   A review of such new technologies can be found in the 2023
community report
\cite{Apresyan:2023frr}, though already this document is somewhat out
of date.

Let me highlight as examples  three directions that are promising areas for
detector R\&D.  The first of these is in tracking, with the use of
Monolithic Active Pixel Sensors (MAPS).    These detectors incorporate
both sensors and readout electronics in wafers with minimal material to
decrease contributions to track pointing resolution from multiple scattering.
The pixel size is less than $25~\mu$; there is no need for bump
bonding.  The wafers can be thinned to less than $50~\mu$, at which
point they are bendable and require only a simple carbon-fiber
support structure.  In principle, such detectors can incorporate
timing and, at the same time, operate at very low power, avoiding the
material of a cryostat\footnote{This goal may be challenging to achieve
  with the high event rates of the Tera-Z program.}.
MAPS are already incorporated in the design of the ALICE
vertex detector upgrade for HL-LHC~\cite{Groettvik:2024onw}, though
this design must
still deal with a high-radiation environment.   The possible
adaptation of this technology to $\ee$ experimentation is still begin explored.

The second is in calorimetry.   From the first studies for
calorimeters for linear $\ee$ collisions, it was envisioned that, in
the $\ee$ environment, one could use particle flow calorimetry
together with a high granularity calorimeter to improve calorimeter
resolution  to $25\%/\sqrt{E}$ and below.   The
calorimeter would be based on a tungsten absorber with small Moli\`ere
radius interleaved with silicon sensors~\cite{Brient:2001fow}.  This
technology has been investigated by the CALICE collaboration; the
current status is described in ~\cite{Poschl:2022lra}.  This strategy
is incorporated in the CMS High-Granularity forward calorimeter for
the HL-LHC upgrade~\cite{Amendola:2025tyi}.

An alternative technology would be dual readout calorimetry, in which
one would collect both scintillation and Cherenkov light signals from
calorimetry modules, allowing separation of the
electromagnetic and non-electromagnetic components of a
hadronic shower~\cite{Lee:2017xss}.   This method is
now undergoing rapid development, with the inclusion of longitudinal
segmentation, timing, and particle flow analysis
strategies~\cite{Akchurin:2022jgd, Eno:2025ltc}.  Up to now,
incorporation of Machine Learning in either technique has not led to
large improvements in resolution, but it is surely just a matter of
time before those ideas have their effect.

The third is in particle ID.  Recently, there have been major advances
in particle ID in the LHC environment from the use of Machine
Learning.  A very influential step was made in the ParticleNet
architecture of Qu and Gouskos~\cite{Qu:2019gqs}.  An application of
this approach  to
full-simulation Higgs factory data by the CEPC group is
reported in \cite{Zhu:2023xpk,Liang:2023wpt}.   These studies are very
promising for the ability to measure $H\to s\bar s$ at  250~GeV.  Here
also, the level of improvement possible in the $\ee$ environment is
still being explored.

This brings up a more general question that I feel is very important.
So far, all of the AI applications to event analysis have been done in
the context of existing designs and hardware, both for LHC and for
Higgs factories.  But to develop a develop a future detector, as AI
gives us more  capabilities, the hardware design should co-evolve with
the analysis software to use the new methods most effectively.  Detectors for
the 2040's will have some level of intelligence on  every sensor.
Does this push us to higher granularity to gain more information
or to lower granularity, with local analysis of the waveforms
produced by the sensors?  How
will we read out highly intelligent sensors?  How will we incorporate
timing information?  How do these solutions interact with  more
traditional restrictions to minimize material and heating?  The
answers to these questions may well  lead us to radically new ideas about
detector design and performance.

This is a fascinating challenge for early-career experimental
physicists.   The people with the best ideas will be the leaders of
collider physics in the Higgs factory era.   These questions need your
attention now.  It is also time to demand serious R\&D funding to
build  prototypes that will validate or challenge your most innovative concepts.

\section{Conclusions}

In this lecture, I have discussed the following points:

\begin{itemize}
  \item The central role of the Higgs boson our understanding of
    particle physics, and the importance of studying the Higgs boson
    at higher levels of precision
  \item The opportunities for the discovery of  signals of physics
    beyond the Standard Model in precision studies of Higgs couplings
    and higher levels of precision in electroweak measurements.
    \item The importance of precision measurements of properties of
      the Higgs boson and the top quark that require $\ee$ CM energies
      well above the $t\bar t$ threshold.
   \item The challenge of constructing Higgs factory detectors of
     ultimate precision to carry out these measurements.
   \end{itemize}

For early-career physicists, the Higgs factory era will be a  time of
adventure and discovery.  Our community needs to bring  this era to
reality on as fast a timeline as possible.

\Acknowledgements

I am grateful to many colleagues for  discussions of the physics
capabilities of $\ee$ Higgs factories over the course of the European
Strategy study.  I thank especially Alain
Blondel,  Jorge de Blas, Ayres Freitas,  Jenny List, Kamal Maayergi,
Matthew McCullough,
Patrick Meade, Roman P\"oschl, Aidan
Robson,  Ariel Schwartzman, Ben Stefanek, Junping Tian, Caterina
Vernieri, Devin Walker, and Filip \.{Z}arnecki.
This work was supported by the US Department of Energy under
                     contract DE--AC02--76SF00515.

\end{document}